\documentclass[useAMS,usenatbib]{mn2e}
\usepackage[]{natbib,amsmath,amssymb,times}
\bibpunct{(}{)}{;}{a}{}{,}

\usepackage[utf8]{inputenc}
\usepackage{eurosym,graphicx,hyperref,epsfig,latexsym}

\usepackage{color}

\renewcommand{\d}{\mathrm{d}}

\newcommand{\ii}{\mathrm{i}}
\newcommand{\bea}{\begin{eqnarray}}
\newcommand{\eea}{\end{eqnarray}}
\newcommand{\be}{\begin{equation}}
\newcommand{\ee}{\end{equation}}
\newcommand{\rund}[1]{\left(#1\right)}
\newcommand{\vc}[1]{\mbox{\boldmath $#1$}}
\newcommand{\dc}{\partial}
\newcommand{\eck}[1]{\left[ #1 \right]}

\def\elabel#1{\label{eq:#1}}

\title[plasma lensing]
{Effects of plasma on gravitational lensing}
\author[Er \& Mao]%
{
Xinzhong Er$^1$\thanks{E-mail: xer@nao.cas.cn}, Shude Mao$^{1,2}$
\\
$^1$National Astronomical Observatories, Chinese Academy of Sciences,
20A Datun Road, Beijing 100012, China\\
$^2$Jodrell Bank Centre for Astrophysics, University of Manchester,
Alan Turing Building, Manchester M13 9PL, UK
}%

\date{Accepted ---; received ---; in original form \today}
\pubyear{2013}

\begin{document}

\maketitle
\begin{abstract}
We study gravitational lensing when plasma surrounds the lens. An
extra deflection angle is induced by the plasma in addition to the
deflection generated by gravity. An inhomogeneous plasma distribution
generates a greater effect than a homogeneous one, and may cause
significant effects to be detected in low frequency radio observations
(a few hundred MHz). In particular, the lensed image positions will be
different between optical and radio observations. The change of
position due to a plasma can reach a few tens of milli-arcsec, which
is readily detectable. One can use the position difference in
different frequencies to estimate the density of plasma in the
lens. The magnification ratios between multiple images are mainly
determined by other properties of the lens, and are only weakly
affected by the plasma. More importantly, we find that the strong
lensing time delay will be affected by the plasma. Estimation of the
Hubble constant from the time delay in low radio frequency observation
may be slightly biased due to plasma in the lens. Unfortunately, the
ionosphere of the Earth strongly affects low frequency radio
observations. Thus our ability to detect the effect depends on how
well we are able to calibrate out the ionosphere.

\end{abstract}

\section{Introduction}

The theory of gravitational lensing is well developed for light
propagation in a vacuum. Deflection angles in a vacuum do not depend
on the photon frequency but only on the mass distribution of the lens. Lensing
provides a way to study mass distribution in the universe, and is
considered to be one of the most powerful tools in cosmology (see
\citealt{2001PhR...340..291B,2010ARA&A..48...87T} for a review). In
the study of cosmological gravitational lensing, several aspects have
been investigated, for instance, the power spectrum of cosmic shear
\citep[e.g.][]{2008A&A...479....9F}, the lensing probability of
the separation of multiple images
\citep[e.g.][]{2001ApJ...549L..25K,2007MNRAS.378..469L}, the
substructure of dark matter haloes \citep[e.g.][]{2013MNRAS.430.3359L}
and time delay between multiple images of strong lens systems
\citep[e.g.][]{2005MNRAS.357..124Y,2010ApJ...712.1378P,2013arXiv1306.4732S}.

It is of interest to study gravitational lensing in a plasma since most
lenses are surrounded by the interstellar or intergalatic
medium. The properties of medium can be characterized by a refractive
index $n$, through which an extra deflection angle can be introduced
into the framework of vacuum lensing theory \citep[see][for more
  detail]{2009GrCo...15...20B, 2010MNRAS.404.1790B,
  2012GrCo...18..117T, 2013PhRvD..87l4009T}. Moreover, the plasma
surrounding the lens is dispersive (the refractive index depends on the
photon frequency) and inhomogeneous. Thus the extra deflection angle
depends on the photon wavelength and the source position.

In general, the extra deflection angle due to plasma is several orders
of magnitude smaller than the gravitational deflection. Only in the
case of a high density plasma and low radio frequency observations (e.g.
a few hundred MHz), is it possible to detect the plasma lensing
effects. Radio images are unaffected by extinction or
microlensing. Thus radio lensing is well developed for studying
substructures, and time delays \citep[e.g.][]{2003ApJ...595..712K,
  2012arXiv1212.2166M}.  With new radio telescopes,
e.g. LOFAR\footnote{http://www.lofar.org} and
SKA\footnote{http://www.skatelescope.org}, high sensitivity,
and high spatial resolution will enable us to measure the weak lensing
shear effect \citep{2013arXiv1303.4650P}. Therefore, the effects due
to plasma in radio lensing may be important in the future.
On the other hand, gravitational lensing preserves the polarization
properties of lensed sources, i.e. fraction and direction. Radiation
of AGNs is polarized \citep[e.g.][]{2005A&A...433..757S}. In fact,
the polarization of some lensed sources has been already measured
\citep{2001ASPC..237...99P}, and can be used to estimate the
magnetic field of the lens using Faraday rotation
\citep{2008arXiv0802.4044N,2013Ap&SS.346..513M}.

In this paper, we will focus on the effects due to plasma in
gravitational lensing. Two simple models of plasma distribution are
employed and two approximations of the plasma lensing equation are
obtained and used. We discuss the effects in estimating weak lensing
shear due to plasma in section 3. The effects on image positions and
time delays are given in section 4.
The cosmology we adopt in this paper is a $\Lambda$CDM model with
parameters based on the results from the PLANCK data
\citep{2013arXiv1303.5076P}: $\Omega_{\Lambda}=0.6825$, $\Omega_{\rm
  m}=0.3175$, a Hubble constant $H_0 = 100 h$ km\,s$^{-1}$\,Mpc$^{-1}$
and $h=0.671$.

\section{Basic Formalism}

The fundamentals of gravitational lensing can be found in
\citet{2001PhR...340..291B}. For its elegance and brevity, we shall
use the complex notation. The thin-lens approximation is adopted,
implying that the lensing mass distribution can be projected onto the
lens plane perpendicular to the line-of-sight. We introduce angular
coordinates $\vc\theta$ with respect to the line-of-sight, and those
on the source plane as $\vc\beta$. The lens equation can be written as
\be
\vc\beta =\vc \theta -\vc \alpha (\vc\theta),
\ee
where $\vc\alpha$ is the deflection angle, and can be calculated from
the lensing potential $\psi$
\be
\vc\alpha =\nabla \psi; \qquad{\rm with} \qquad
\nabla = {\dc \over \dc \theta_1} + \ii {\dc \over \dc \theta_2}.
\elabel{alpha}
\ee
The lensing potential is determined by the dimensionless projected surface-mass density (lensing convergence),
\bea
\psi(\vc\theta) &=& \frac{1}{\pi}\int_{{\cal R}^2} \d^2\theta'\kappa(\vc\theta')\;
    {\rm ln}|\vc\theta-\vc\theta'|\;;\\
\kappa(\vc\theta) &=& \Sigma(\vc\theta)/\Sigma_{\rm cr},\;\;\; {\rm where} \;\;\;
\Sigma_{\rm cr} = \frac{c^2}{4\pi G} \frac{D_{\rm s}}{D_{\rm d} D_{\rm ds}}\;
\eea
is the critical surface mass density depending on the angular-diameter
distances $D_{\rm s}$, $D_{\rm d}$ and $D_{\rm ds}$ from the observer
to the source, the observer to the lens, and the lens to the source,
respectively.  $\Sigma(\vc\theta)$ is the projected surface-mass
density of the lens.
To the lowest order, image distortions caused by gravitational lensing are
described by the complex shear
\be
\gamma = \frac{1}{2}\left(\partial_1^2\psi-\partial_2^2\psi\right)
+ {\rm i}\partial_1\partial_2\psi\;,
\elabel{shear}
\ee
which transforms a round source into an elliptical shape.
The magnification for a point source is given by
\be
\mu = \dfrac{1}{(1-\kappa)^2 \,-\,|\gamma|^2}.
\elabel{anamu}
\ee

\subsection{Gravitational lensing in plasma}
In presence of a medium around the lens, the deflection angle will be
slightly changed. In the weak field approximation, the
lensing deflection angle in plasma is given by
\citep{2009GrCo...15...20B,2010MNRAS.404.1790B} as
\bea
\vc\alpha_{pl} ={\vc \alpha }\;
&+& {\vc \alpha \over 2} \rund{ \int_0^{\infty} {1\over 1-\omega_e^2/\omega^2}
{b^2\over(b^2+x_3^2)^{3/2}} \d x_3 -1} \nonumber\\
&+& {\omega_e^2 \over \omega^2} {1\over N_e} \int_0^{\infty}
{\dc N_e \over \dc b} \d x_3
\elabel{angleplasma}
\eea
where the first term is the vacuum gravitational deflection; the
second term is the additional deflection due to the presence of
a homogeneous plasma; the third term is the deflection due to the plasma
inhomogeneity (the refraction);
$\omega$ is the photon frequency; and $\omega_e$ is the electron
plasma frequency. The plasma frequency is determined by the density
and mass of the ionized gas, i.e. $\omega_e^2= 4 \pi e^2 N_e /m_e$.
The refractive index of the plasma is given by
$n=1-\omega_e^2/\omega^2$, and thus the phase speed of light in the
plasma is $v=c/n$. $N_e$ is the number density of electron, $b$ is the
impact parameter, and $x_3$ is the coordinate along the line of
sight. The formula (Eq.~\ref{eq:angleplasma}) is valid only for
$\omega>\omega_e$, since light waves with $\omega<\omega_e$ do not
propagate in the plasma. In the limit of $\omega>>\omega_e$, this
formula reduces to the vacuum case. The presence of plasma changes the
deflection angle with the difference from the vacuum case being
strongest for long wavelengths, as $\omega$ approaches $\omega_e$.
The homogeneous plasma (second term in Eq.~\ref{eq:angleplasma})
increases the deflection angle of gravitational lensing. The density
of plasma in galaxies usually decreases with radius, thus the
refraction deflection (third term in Eq.~\ref{eq:angleplasma}) is
opposite to the gravitational deflection, since the refractive index
of plasma is small than $1$.

In general, the photon frequency is much larger than $\omega_e$. Only
in the case of radio wavebands and high density plasma, does the
difference from the vacuum case reach a few percent. In this paper, we
study lensing with low frequency radio observations in order to assess
the observational signature of plasma. Two plasma distribution models,
typical of spiral and elliptical lensing galaxies are adopted in this
work. We discuss them in turn.

In model 1, appropriate to a spiral lensing galaxy, we assume that the
number density of electrons is
$10$~cm$^{-3}$ \citep{2010ApJ...710L..44G}, which corresponds to a
plasma frequency of $\sim1.8\times10^5$~Rad/s. The distribution of plasma
around a real lens is complex, so we use a circular symmetric form
as an approximation. This is consistent with observations from a
nearby galaxy, $M51$ \citep{2010ApJ...710L..44G}. Taking
\be
N_e(r)=N_0 {\rm e}^{-r/r_0}
\ee
where $N_0$ is the central density of plasma, $r$ is the radial
distance from the centre of the galaxy, and $r_0$ is the scale radius
of the central region. For $M51$, $N_0=10\pm1$ cm$^{-3}$, and
$r_0=10\pm 1\,$kpc. The deflection can thus be simplified to
\be
\vc\alpha_{pl} = \vc \alpha (1+{\omega_e^2\over 2 \omega^2})
- {\omega_e^2 \over \omega^2}\, {b\over r_0} F(b) \hat\theta,
\elabel{plasmafull}
\ee
where $\hat\theta$ is the unit vector of coordinates,
and $F(b)$ is an integral given in the Appendix.

In general, the observational frequency is several orders of magnitude
higher than
the plasma frequency ($\omega_e^2/\omega^2<10^{-6}$). It is only
possible to identify the plasma lensing effect in radio observations
at low frequencies (e.g. a few hundreds MHz). From algebra, one can see
that the additional deflection due to the homogeneous plasma is much
smaller than that due to the gradient of the plasma. Thus we only consider
the additional deflection due to an inhomogeneous plasma in this
paper, and use a simplified lens equation with plasma
\be
\vc\alpha_{pl} (\vc\theta)=\vc \alpha(\vc \theta)
- 0.031\,{\vc\theta}\,f(\theta),
\elabel{plangle}
\ee
where $\vc\theta$ is the angular coordinates vector from the centre of
the galaxy. More detail about the derivation of Eq.~(\ref{eq:plangle})
and the integral function $f(\theta)$ can be found in Appendix A. Note
that Eq.~(\ref{eq:plangle}) is only valid for our model of plasma
distribution and our observational frequency ($375$ MHz). For general
cases, Eq.~(\ref{eq:plasmafull}) must be used instead.

In the second model of plasma in a lens galaxy, we use a more
conservative estimate of the electron number density appropriate for
an elliptical galaxy. It is given by \citep{2003ARA&A..41..191M}
\be
N_e(r) = N_0 (r/r_0)^{-1.25},
\ee
where $N_0=0.1$cm$^{-3}$ is again the central density of plasma, and
$r_0=10$~kpc is the scale radius. We also have a simplified lens
equation for the second model
\be
\vc\alpha_{pl} (\vc\theta) = \vc\alpha(\vc\theta) -
0.0006 \rund{\theta_0\over \theta}^{5/4} \hat\theta
\elabel{plangle2}
\ee
where $\theta_0=1.6$ arcsec is the projected scale radius, and $\hat
\theta$ is the unit vector. More detail is contained in Appendix B.
In Fig.\ref{fig:compala}, we compare the extra deflection angle due to
two plasma models. As one might expect, the deflection due to the first plasma
model is larger than that of the second model. The main reason is that the
electron density in model 2 is smaller than that in model 1.  Both
kinds of galaxies (spiral and elliptical) have been found as lens
galaxies in real observations. The plasma distributions that we use
here are two typical models for electron density: the spiral
galaxy is an upper limit, while the elliptical galaxy is a lower limit.

\begin{figure}
\centerline{\scalebox{1.0}
{\includegraphics[width=7cm, height=5cm]{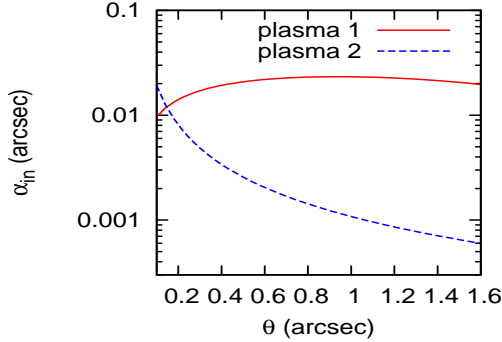}}}
\caption{The extra deflection due to the inhomogeneous distribution of
  plasma. The solid red line represents the deflection angle of the first
  plasma model (Eq.~\ref{eq:plangle}), while the blue dashed line represents
  the second model (Eq.~\ref{eq:plangle2}).}
\label{fig:compala}
\end{figure}

Throughout this paper, we adopt the same mock lens galaxy with both plasma
models. The lens redshift is $0.5$, and the source redshift is $2.0$.
We use a Singular Isothermal Ellipsoid (SIE) for the lens mass model
with a velocity dispersion of $\sigma_v=140$~km/s (which corresponds to
$\theta_{\rm E}=0.40$ arcsec at this redshift), and ellipticity
$\epsilon=0.3$.

\section{Effect on shear}

The additional deflection angle due to plasma decreases with increasing angular
separation $\theta$. The deflections are generally small and can be neglected
with large $\theta$, e.g. $\theta>10$ arcsec. Moreover, the plasma models are no
longer valid at large radius. We thus only consider
small angular separations in this paper.

\begin{figure}
\centerline{\scalebox{1.0}{
\includegraphics[width=7cm, height=7cm]{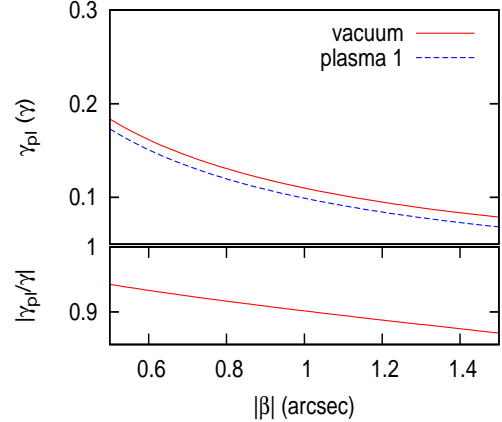}}}
\caption{Estimates of weak lensing shear using ellipticity for the plasma
  model 1. In the top panel, the solid (dashed) line represents the
  estimated shear as a function of the source position in vacuum (plasma)
  lensing. In the bottom panel, the line shows the ratio of estimated
  shear of the plasma lensing over vacuum lensing as a function of
  source position.}
\label{fig:shear}
\end{figure}

\begin{figure}
\centerline{\scalebox{1.0}{
\includegraphics[width=7cm, height=7cm]{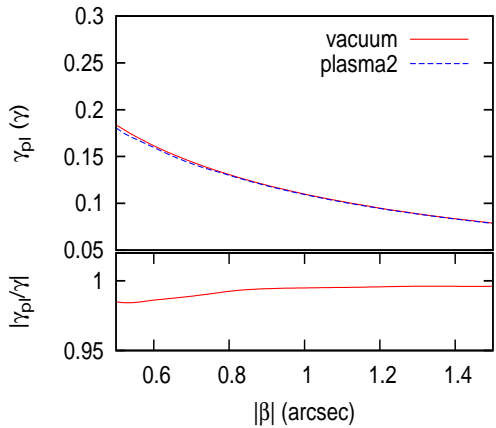}}}
\caption{Same as Fig.~\ref{fig:shear} but for plasma model 2.}
\label{fig:shear2}
\end{figure}

We perform numerical simulations to test the changes in shear due to plasma.
A circular-symmetric light profile is used for the source image
\be
I(\beta)=I_e {\rm exp}\eck{-7.67((\beta/\beta_e)^{1/4}-1)},
\elabel{lightprofile}
\ee
where $I_e=1$ in arbitrary units, and $\beta_e=0.05$ arcsec. We
move the source along the $\beta_2$ direction from $(0.2,0)$ to
$(0.2,2)$ in order to generate a series of images. The lens equations in
vacuum and with plasma (Eq.~\ref{eq:plangle}) are used separately to
generate two sets of lensed images. We use the second order brightness
moments as an estimate of the ellipticity (shear) \citep{1995ApJ...449..460K}
\be
\gamma={Q_2\over 2 Q_0} = {\int \d^2 \theta\, I(\theta)\, \theta^2
\over 2 \int \d^2\theta\, I(\theta)\, \theta\theta^*}.
\ee
We compare the estimated ellipticity between the two sets of images
(Figs.~\ref{fig:shear} and \ref{fig:shear2}). We can see that the
presence of plasma increases the ellipticity of lensed images. In the
case of very low frequency band observations and the high density plasma
model 1, the estimated deviation can reach $\sim5-10\%$.
Although the fractional changes become slightly larger at large
radius, the shear decreases more rapidly. The effect of plasma at
large radius is in fact more difficult to observe. In the plasma model
2, there is only about $\sim1\%$ difference.

Moreover, as we will see in the next section, the image
positions are also slightly changed due to plasma. The combined
effects (shear increasing and position shift) will change the estimated
shear power spectrum, and cause slight systematics in the study of
cosmic shear.


\section{Effects on strong lensing}

\subsection{Image positions and magnifications}
\begin{figure}
\centerline{\scalebox{1.0}
{\includegraphics[width=8cm, height=6cm]{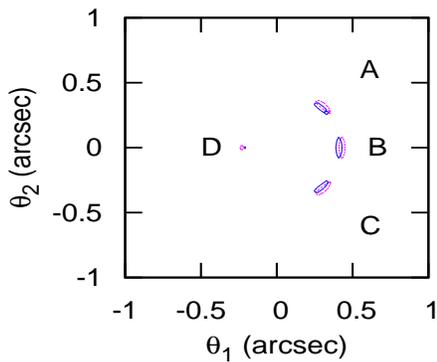}}}
\caption{ Two strong lensing multiple images: the purple images are
  the ones generated by vacuum lens equation, while the blue images
  are those with plasma (model 1).
  The lens is at the origin of the figure. The images are labelled
  from A to D.}
\label{fig:images}
\end{figure}

Lensing magnification in presence of plasma is more complicated than
shear since it is not linearly related to the deflection
angle. Moreover, the absorption and scintillation of plasma will
change the luminosity of the images, and the precise effect is
wavelength dependent \citep[see e.g.][]{2004ARA&A..42..275S}. We will
not consider the absorption effect since it is not significant in
radio observations. The magnifications of the multiple images are
changed differently by plasma.

With the same lensing properties as in the previous section, we create
a mock multiple image strong lensing system. The source is placed at
$(0.1,0.0)$ with a similar bright profile as in
Eq.~(\ref{eq:lightprofile}).  We use a more compact source
($\beta_e=0.01$ arcsec) in order to isolate the magnification
effect. In Fig.~\ref{fig:images}, we can see that four images
(labelled A to D) are generated by lensing. Similar strong lensing
systems have been found in the Universe
\citep[e.g.][]{2002A&A...388..373B}. The magnification $\mu$ and
magnification ratio of the multiple images are given in
Table~\ref{table1}. The images generated by plasma lensing (blue in
Fig.~\ref{fig:images}) have slight smaller image separations.
We can fit the plasma lensing images with the vacuum lens
equation. The strong lens modelling is complicated and beyond the scope
of this paper, and one can find more detail from
e.g. \citet{2009ApJ...691..277S} and \citet{2009MNRAS.392..945V}. We
evaluate only two parameters, the source position and the Einstein
radius, and assume accurate knowledge of the other parameters.  We
obtain a slightly different source position and Einstein radius,
$\beta=(0.095,0)$, $\theta_{\rm E}=0.382$ arcsec, which is about $5\%$
smaller than the input value. For the plasma model 2, we obtain
$\beta=(0.099,0)$, $\theta_{\rm E}=0.401$ arcsec, which is only $1\%$
different from the true model. The magnifications predicted by the
models (also given in Table~\ref{table1}) have not significantly
changed. The magnification cannot be observed directly since we do not
know the intrinsic flux of the source. The flux ratio of multiple
images can be directly compared between observations and model
predictions. However, the magnification ratios are strongly affected by
the lens properties, such as substructures
\citep{1998MNRAS.295..587M,2012MNRAS.421.2553X}. The effects due to
plasma may be not as great as those due to substructures, and thus
plasma has no important role in explaining the flux ratio anomalies.
\begin{center}
\begin{table*}
\begin{tabular}{|c|c|c|c||c|c|c|c|c|c|c|}
\hline \hline
       &$\theta_A$ &$\theta_B$ &$\theta_C$ &$\theta_D$ &$\mu_A$ &$\mu_B$ &$\mu_C$ &$\mu_D$ &$\mu_A/\mu_B$ &$\mu_A/\mu_D$\\ \hline
vacuum &$(0.304,0.322)$ &$(0.432,0)$ &$(0.304,-0.322)$ &$(-0.23,0)$ &$4.39$ &$5.13$ &$4.39$ &$0.811$ &$0.856$ &$6.33$\\ \hline
plasma 1 &$(0.290,0.310)$ &$(0.413,0)$ &$(0.290,-0.310)$ &$(-0.217,0)$ &$4.13$ &$4.83$ &$4.13$ &$0.734$ &$0.855$ &$6.57$\\ \hline
fitting 1 & & & & &$4.37$ &$5.09$ &$4.37$ &$0.816$ &$0.860$ &$6.23$\\ \hline
plasma 2 &$(0.302,0.322)$ &$(0.429,0)$ &$(0.302,-0.322)$ &$(-0.225,0)$ &$4.40$ &$5.15$ &$4.40$ &$0.821$ &$0.854$ &$6.28$\\ \hline
fitting 2 & & & & &$4.33$ &$4.98$ &$4.33$ &$0.823$ &$0.868$ &$6.05$\\ \hline
\hline
\end{tabular}
\caption{\label{table1} The first four columns are the positions of
  lensed images (in units of arcsec). The other columns are
  magnifications and magnification ratios of the strong lensing
  multiple images with/without plasma. The third and fifth lines show
  predictions of magnification from fitting the vacuum to plasma model
  (see section 4.1 for more details). The positions and labels (A-D)
  are shown in the Fig.~\ref{fig:images}. }
\end{table*}
\end{center}

\subsection{Time delay}
For a given source position, the excess time-delay surface as a function of
position in the image plane is given by
\bea
t(\vc\theta) &=& {D_d D_s \over D_{ds}} {1+z_d \over c}
\eck{{1\over 2}|\vc\theta-\vc\beta|^2 - \psi(\vc\theta)}\\
&\propto& {1\over H_0}
\eck{{1\over 2}|\vc\theta-\vc\beta|^2 - \psi(\vc\theta)}
\elabel{timedelay}
\eea
In the presence of plasma, the image position, $\theta_{pl}$, will be
different.  In reality, an extra effect should also come into the
geometrical term ${1\over 2}|\vc\theta-\vc\beta|^2$ since the speed of
light is different from $c$ in plasma. However, the refractive index
is very close to $1$, even in the case of high density interstellar
media and low frequency observations. The effect on the geometrical
term is below the order of $10^{-5}$, and cannot be determined by
current observational techniques. It will not be considered further
here.

We calculate the time delay between multiple images, e.g.,~$\Delta
t_{AB}=|t(\theta_A)-t(\theta_B)|$. The results of using vacuum and
plasma lensing equations are given in Table~\ref{table2}. The
difference in the time delay is about $0.2-1\%$. The bottom line in
Table~\ref{table2} shows the time delay results calculated from
fitting the model in the previous section. One can see that the
difference is slightly larger.

By using the image configuration, one can model the mass distribution
of the lens to determine the lens potential $\psi(\theta)$ and the
unlensed source position $\vc\beta$. Lens systems with time delays can
thus be used to study the cosmological parameters, e.g. Hubble
constant (\citealt{1964MNRAS.128..307R}, one can also read more detail
from the COSMOGRAIL project, e.g.,
\citealt{2011A&A...536A..53C,2013A&A...556A..22T}). One can also see
that there is a degeneracy between the cosmology and the lens
potential, in that the mass-sheet degeneracy will cause an
underestimate of the Hubble constant
\citep[e.g.][]{2013arXiv1306.0901S}.
We use the mock data generated from plasma lensing and use the vacuum
lensing equation to reconstruct the mass distribution, and then
calculate the Hubble constant. Other cosmological parameters, such as
$\Omega_\Lambda$, are unchanged in order to isolate the effect of
plasma lensing in determining Hubble constant. The lensing model is
fitted to the multiple image positions, given in the previous
section. The Hubble constant can be estimated from two time delays:
$h=0.628$ from $\Delta t_{AB}$ and $h=0.613$ from $\Delta
t_{AD}$. Both results give significant underestimates at $\sim9\%$.
For plasma model 2, we obtain $h=0.633$ from $\Delta t_{AB}$ and
$h=0.668$ from $\Delta t_{AD}$.

Some analyses use optical observations for image positions and radio
observations for the time delay measurement
\citep{2002ApJ...581..823F,2010ApJ...711..201S}. In this case, the
image positions are not modified by the plasma. Thus the lensing mass
model is unbiased, while the time delay is still changed by the
plasma. Such an approach will however lead to a slight underestimate
for the Hubble constant. Using our mock lensing data, we obtain
$h=0.673$ from $\Delta t_{AB}$ and $h=0.672$ from $\Delta t_{AD}$. The
approach can cause an underestimate of up to $1\%$ in the Hubble
constant. In the second plasma model, there is no significant bias in the
estimates.
\begin{center}
\begin{table}
\begin{tabular}{|c|c|c|c|}
\hline\hline
 &$\Delta t_{AB}$  & $\Delta t_{AC}$  &$\Delta t_{AD}$\\ \hline
vacuum &0.454  &0  &6.21  \\ \hline
plasma 1 &0.453  &0  &6.20  \\ \hline
fitting 1 &0.424  &0  &5.66  \\ \hline
plasma 2 &0.454 &0 &6.21 \\ \hline
fitting 2 &0.481 &0 &6.18 \\ \hline
\hline
\end{tabular}
\caption{\label{table2} Time delay in units of day between lensed
  multiple images in vacuum, two plasma lensings and two model
  predictions. The image positions and labels are shown in
  Fig.~\ref{fig:images}. }
\end{table}
\end{center}

\section{Polarization and magnetic field}
When a light ray passes through a plasma in the presence of a magnetic field,
the polarization vector rotates due to the magnetic field.
The rotation angle of the plane of polarization is given by
\citep{2004PhRvD..69h7501S,2013Ap&SS.346..513M}
\be
\phi = {e^3 \lambda^2\over 2 \pi m_e^2 c^4} \int_L B_{\parallel}(l)\,N_e(l)\,\d l,
\ee
where $\lambda$ is the wavelength of the radiation as seen by
the absorber medium, $B_{\parallel}$ is the line of sight component of
the magnetic field, and the integral is over the path length through
the intervening absorbers. $m_e$ is the mass of the electron, and
$e$ is the elementary charge.
In multi-wavelength observations of polarization, the different
rotation angles are determined from the Faraday Rotation Measurement (RM)
\be
RM={\Delta \phi\over \lambda_{\rm obs}^2} = {e^3 \over 2\pi m_e^2c^4}
\int B_{\parallel}(l) N_e(l) \rund{\lambda \over \lambda_{\rm obs}}^2 \d l,
\ee
where $\Delta \phi=\phi(\lambda)-\phi(0)$. In principle, from the RM one
can estimate the average magnetic field of the lens
\be
\langle B_{\parallel} \rangle ={\int N_e(l) B_{\parallel}(l) \d l
  \over \int N_e(l) \d l},
\ee
where the denominator is the electron column density.

In reality measurement of polarization is difficult.  Some results
with a few percent polarization have been carried out
\citep{2007MNRAS.380..162J,2011MNRAS.413..132B}. The contamination
along the line of sight, mainly from the Milky Way, is large. In
multiple image lensing systems, the different images propagate along
different lines of sight with different magnetic fields and will
experience different RMs. We can use the differences to estimate the
variation of magnetic field around the lens
\citep{2001ApJ...562..649K}. This estimate is independent of
contamination along the line of sight since the angular separation is
small, and the variation on such scales can be neglected. The time
delay effect will change the RM due to source variations. Thus one has
to calibrate the RM at the same emission time.
We do not attempt to model the polarization within this paper since
there is not sufficient observation result.

\section{Summary and Discussion}

In this paper, we have studied gravitational lensing in the presence
of plasma in the lens galaxy. Usually the lens galaxy/cluster is
surrounded by plasma, and an extra deflection angle is caused by the
plasma, especially by plasma inhomogeneity. By adopting a
plasma distribution model in the lens galaxy, we have obtained a
plasma lensing equation for a given range and observational
frequency. The presence of plasma can cause several changes in
lensing: the positions and magnifications of the lensed images, shear,
and the time delay between the multiple images in strong lensing are
all altered.

In general, as most lenses are elliptical galaxies,
the electron number density in the galaxy is probably as
low as the second model we used in this study.
Thus the effect due to plasma in lensing may not
be significant. However, galaxies at high redshift may be
different with denser ionized gas. The inclusion of plasma lensing
provides a way to estimate the density of plasma in the lens. The
greater the difference between image positions in different wave bands, the
higher the density of plasma.

The effects due to plasma are only significantly observable in very
low frequency observations. Radio observations using VLBI can reach
very low frequencies ($\sim100$ MHz), and very high spatial resolution
(milli-arcsec). The plasma frequency in a high electron density region
can reach $\sim10^6$Hz. Under such conditions, the signatures due to
plasma in lensing can reach a few percent. Magnification ratios
between multiple images are mainly determined by the overall mass
distribution and any inhomogeneous mass distribution within the lens,
i.e. substructures. Therefore, the effect due to plasma on
magnification ratios will be hard to detect. On the other hand, the
impact of plasma on image positions and time delays may be
discernible. It will be possible to use the image position differences
between radio and optical observations to estimate the plasma
density. Moreover, the effects on the time delay will bias estimates
of the Hubble constant determined using low radio frequency. One
should take the plasma effect into account in estimating the Hubble
constant from low frequency radio observations.

However, intrinsic source images are different at different
wavelengths. Especially in radio observations, the source size is
larger than that in the optical. This effect will cause differential
magnifications \citep{2013ApJ...770..110E}, and thus introduce more
difficulty in the reconstruction of the lens and source. Nevertheless, the
observations from different wavelengths provide more information on
both the lens and source.

Due to difficulties in the measurement of polarization and the
contamination along the line of sight, little knowledge of lens magnetic
fields has been obtained. With the help of new radio
telescopes, such as LOFAR and SKA, it may be possible, using a large lens
sample, to place constraints on the magnetic fields of lens galaxies or
clusters.

One potential difficulty is that the ionosphere of the Earth affects
radio transmission. It introduces systematics into the image
positions, which drift around the sky with time. The state of the
ionospheric plasma is difficult to describe, since it is strongly
dependent on the activity of the Sun. For instance, the local winter
hemisphere is tipped away from the Sun, and thus the ionosphere has less
influence. The typical electron column at night
\footnote{http://iono.jpl.nasa.gov/latest\_rti\_global.html} is
$\sim10^{17}$~m$^{-2}$.
The electron column density in the galaxy is larger
$\sim10^{25}$~m$^{-2}$ (under the assumption of $n_e=0.1$~cm$^{-3}$
and the thickness of the plasma is $10$~kpc). However, in order to
obtain precise image positions, we have to use long baseline
interferometry. The states of the ionosphere at different telescope
sites are different \citep{2006AAS...209.8506C}. Calibration is
essential but very difficult. Space radio telescopes, like
Spekr-M\footnote{http://www.russianspaceweb.com/spektr$\_$m.html}, may
provide more information for calibration and observing lensing
images. However, image stabilities at the milli-arcsec level are
required by our study, but remain to be demonstrated in reality.


\section*{Acknowledgments}
We thank the referee for important comments, especially for pointing
out a major mistake in the draft. We thank Neal Jackson, Richard
J. Long, Lijun Gou and Olaf Wucknitz for useful comments on the
draft. X.E. is supported by NSFC grant No.11203029. S.M. is supported
by the Chinese Academy of Sciences and the National Astronomical
Observatories of China.

\appendix
\section{Deflection angle integral of model 1}
In this section, we will neglect the effect of homogeneous
plasma. The deflection angle caused by the plasma gradient is given
by
\be
\alpha_{in}=\int_0^{\infty} {1\over \omega^2} K_e {\dc N_e \over \dc b} \d x_3,
\ee
where $K_e=4\pi e^2/m_e$. The electron density we adopt in this
paper is measured from the nearby galaxy M51 \citep{2010ApJ...710L..44G}
\be
N_e= N_0 {\rm e}^{-r/r_0},
\ee
where $N_0=10$~cm$^{-3}$ is a central value of electron density and
$r_0$ is a scale length, which takes the value of $10$~kpc. $r$ is
the spherical radius $r=\sqrt{x_3^2+b^2}$. The impact parameter $b$ is
approximately $b=\theta D_d$. The deflection angle can be written as
\be
\vc\alpha_{in}(\vc\theta) = -{\omega_0^2 \over \omega^2} {\vc\theta\over \theta_0}
F(\theta) \qquad {\rm with}\quad F(\theta)=\int_0^{\infty}
\dfrac{{\rm e}^{-\theta/\theta_0\sqrt{1+x^2}}}{\sqrt{1+x^2}} \d x,
\ee
where $\omega_0^2=K_eN_0$, and $\theta_0=r_0/D_d$($\approx1.6$ arcsec in our
case). The integral function $F(\theta)$ can be approximated
by a fitting function
\be
f(\theta)=3.6\,\theta^{-0.22} - 2.85,
\elabel{fbapprox}
\ee
where $\theta$ is in units of arcsec and $f(\theta)$ is dimensionless.
One should note that Eq.~(\ref{eq:fbapprox}) is only valid for the
plasma distribution used in this work, i.e. $\theta_0=1.6$ arcsec and is
in the range of $[0.3,1.5]$ arcsec. Further, we assume a radio
frequency ($375$MHz). The total
deflection angle by gravity and plasma is calculated by
\be
\vc\alpha_{pl}=\vc \alpha - 0.031\,{\vc\theta}\,f(\theta),
\elabel{applenseq}
\ee
where the last term is given in units of arcsec.
\begin{figure}
\centerline{\scalebox{1.0}
{\includegraphics[width=7cm, height=6cm]{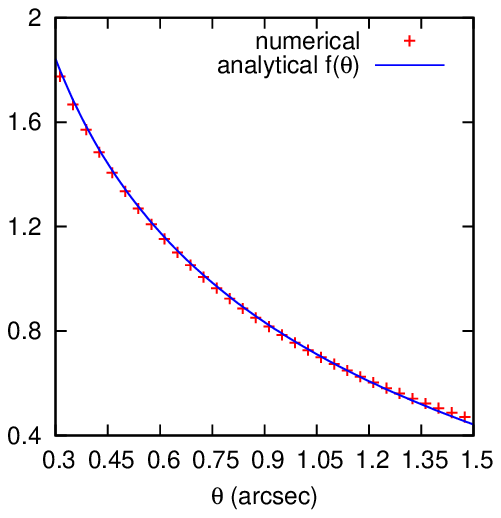}}}
\caption{The integral function $F(\theta)$ (plus points) and its
  approximation $f(\theta)$ (solid line, Eq.~\ref{eq:fbapprox}).}
\label{fig:fbx}
\end{figure}

\section{Deflection angle of model 2}
In the second plasma model, the plasma density is given by
\citet{2003ARA&A..41..191M} and is appropriate for a typical elliptical galaxy
\be
N_e= N_0 (r/r_0)^{-1.25},
\ee
where we have $N_0=0.1$ cm$^{-3}$ and $r_0=10$ kpc.  The deflection
angle due to this inhomogeneous distribution of plasma is given by
\be
\alpha_{in}= -{\omega_0^2\over \omega^2} \rund{r_0 \over b}^{5/4}
\sqrt{\pi} \;{\Gamma(9/8)\over \Gamma(5/8)}.
\ee
For our lensing configuration and observation frequency ($375$ MHz), the total
deflection angle by gravity and plasma is calculated as
\be
\vc \alpha_{pl} = \vc \alpha - 0.0006 \rund{\theta_0\over\theta}^{1.25}\hat\theta,
\ee
where $\theta_0=1.6$ arcsec and $\hat \theta$ is the unit vector.

\bibliographystyle{aa}
\bibliography{../../../bib/refbooks,../../../bib/lens,../../../bib/refcos,../../../bib/qso,../../../bib/stronglens,../../../bib/bhlens,../../../bib/galaxy}

\begin{thebibliography}{42}
\expandafter\ifx\csname natexlab\endcsname\relax\def\natexlab#1{#1}\fi

\bibitem[{{Bartelmann} \& {Schneider}(2001)}]{2001PhR...340..291B}
{Bartelmann}, M. \& {Schneider}, P. 2001, \physrep, 340, 291

\bibitem[{{Battye} {et~al.}(2011){Battye}, {Browne}, {Peel}, {Jackson}, \&
  {Dickinson}}]{2011MNRAS.413..132B}
{Battye}, R.~A., {Browne}, I.~W.~A., {Peel}, M.~W., {Jackson}, N.~J., \&
  {Dickinson}, C. 2011, \mnras, 413, 132

\bibitem[{{Bisnovatyi-Kogan} \& {Tsupko}(2009)}]{2009GrCo...15...20B}
{Bisnovatyi-Kogan}, G.~S. \& {Tsupko}, O.~Y. 2009, Gravitation and Cosmology,
  15, 20

\bibitem[{{Bisnovatyi-Kogan} \& {Tsupko}(2010)}]{2010MNRAS.404.1790B}
{Bisnovatyi-Kogan}, G.~S. \& {Tsupko}, O.~Y. 2010, \mnras, 404, 1790

\bibitem[{{Brada{\v c}} {et~al.}(2002){Brada{\v c}}, {Schneider}, {Steinmetz},
  {Lombardi}, {King}, \& {Porcas}}]{2002A&A...388..373B}
{Brada{\v c}}, M., {Schneider}, P., {Steinmetz}, M., {et~al.} 2002, \aap, 388,
  373

\bibitem[{{Cotton} \& {Uson}(2006)}]{2006AAS...209.8506C}
{Cotton}, Jr., W.~D. \& {Uson}, J. 2006, in Bulletin of the American
  Astronomical Society, Vol.~38, American Astronomical Society Meeting
  Abstracts, 1015

\bibitem[{{Courbin} {et~al.}(2011){Courbin}, {Chantry}, {Revaz}, {Sluse},
  {Faure}, {Tewes}, {Eulaers}, {Koleva}, {Asfandiyarov}, {Dye}, {Magain}, {van
  Winckel}, {Coles}, {Saha}, {Ibrahimov}, \& {Meylan}}]{2011A&A...536A..53C}
{Courbin}, F., {Chantry}, V., {Revaz}, Y., {et~al.} 2011, \aap, 536, A53

\bibitem[{{Er} {et~al.}(2013){Er}, {Ge}, \& {Mao}}]{2013ApJ...770..110E}
{Er}, X., {Ge}, J., \& {Mao}, S. 2013, \apj, 770, 110

\bibitem[{{Fassnacht} {et~al.}(2002){Fassnacht}, {Xanthopoulos}, {Koopmans}, \&
  {Rusin}}]{2002ApJ...581..823F}
{Fassnacht}, C.~D., {Xanthopoulos}, E., {Koopmans}, L.~V.~E., \& {Rusin}, D.
  2002, \apj, 581, 823

\bibitem[{{Fu} {et~al.}(2008){Fu}, {Semboloni}, {Hoekstra}, {Kilbinger}, {van
  Waerbeke}, {Tereno}, {Mellier}, {Heymans}, {Coupon}, {Benabed}, {Benjamin},
  {Bertin}, {Dor{\'e}}, {Hudson}, {Ilbert}, {Maoli}, {Marmo}, {McCracken}, \&
  {M{\'e}nard}}]{2008A&A...479....9F}
{Fu}, L., {Semboloni}, E., {Hoekstra}, H., {et~al.} 2008, \aap, 479, 9

\bibitem[{{Guti{\'e}rrez} \& {Beckman}(2010)}]{2010ApJ...710L..44G}
{Guti{\'e}rrez}, L. \& {Beckman}, J.~E. 2010, \apjl, 710, L44

\bibitem[{{Joshi} {et~al.}(2007){Joshi}, {Battye}, {Browne}, {Jackson},
  {Muxlow}, \& {Wilkinson}}]{2007MNRAS.380..162J}
{Joshi}, S.~A., {Battye}, R.~A., {Browne}, I.~W.~A., {et~al.} 2007, \mnras,
  380, 162

\bibitem[{{Kaiser} {et~al.}(1995){Kaiser}, {Squires}, \&
  {Broadhurst}}]{1995ApJ...449..460K}
{Kaiser}, N., {Squires}, G., \& {Broadhurst}, T. 1995, \apj, 449, 460

\bibitem[{{Keeton} \& {Madau}(2001)}]{2001ApJ...549L..25K}
{Keeton}, C.~R. \& {Madau}, P. 2001, \apjl, 549, L25

\bibitem[{{Kemball} {et~al.}(2001){Kemball}, {Patnaik}, \&
  {Porcas}}]{2001ApJ...562..649K}
{Kemball}, A.~J., {Patnaik}, A.~R., \& {Porcas}, R.~W. 2001, \apj, 562, 649

\bibitem[{{Koopmans} {et~al.}(2003){Koopmans}, {Biggs}, {Blandford}, {Browne},
  {Jackson}, {Mao}, {Wilkinson}, {de Bruyn}, \&
  {Wambsganss}}]{2003ApJ...595..712K}
{Koopmans}, L.~V.~E., {Biggs}, A., {Blandford}, R.~D., {et~al.} 2003, \apj,
  595, 712

\bibitem[{{Li} {et~al.}(2007){Li}, {Mao}, {Jing}, {Lin}, \&
  {Oguri}}]{2007MNRAS.378..469L}
{Li}, G.-L., {Mao}, S., {Jing}, Y.~P., {Lin}, W.~P., \& {Oguri}, M. 2007,
  \mnras, 378, 469

\bibitem[{{Li} {et~al.}(2013){Li}, {Mo}, {Fan}, {Yang}, \&
  {Bosch}}]{2013MNRAS.430.3359L}
{Li}, R., {Mo}, H.~J., {Fan}, Z., {Yang}, X., \& {Bosch}, F.~C.~v.~d. 2013,
  \mnras, 430, 3359

\bibitem[{{MacLeod} {et~al.}(2012){MacLeod}, {Jones}, {Agol}, \&
  {Kochanek}}]{2012arXiv1212.2166M}
{MacLeod}, C.~L., {Jones}, R., {Agol}, E., \& {Kochanek}, C.~S. 2012, ArXiv:
  1212.2166

\bibitem[{{Mao} \& {Schneider}(1998)}]{1998MNRAS.295..587M}
{Mao}, S. \& {Schneider}, P. 1998, \mnras, 295, 587

\bibitem[{{Mathews} \& {Brighenti}(2003)}]{2003ARA&A..41..191M}
{Mathews}, W.~G. \& {Brighenti}, F. 2003, \araa, 41, 191

\bibitem[{{Morozova} {et~al.}(2013){Morozova}, {Ahmedov}, \&
  {Tursunov}}]{2013Ap&SS.346..513M}
{Morozova}, V.~S., {Ahmedov}, B.~J., \& {Tursunov}, A.~A. 2013, \apss, 346, 513

\bibitem[{{Narasimha} \& {Chitre}(2008)}]{2008arXiv0802.4044N}
{Narasimha}, D. \& {Chitre}, S.~M. 2008, ArXiv:0802.4044

\bibitem[{{Paraficz} \& {Hjorth}(2010)}]{2010ApJ...712.1378P}
{Paraficz}, D. \& {Hjorth}, J. 2010, \apj, 712, 1378

\bibitem[{{Patel} {et~al.}(2013){Patel}, {Abdalla}, {Bacon}, {Rowe}, {Smirnov},
  \& {Beswick}}]{2013arXiv1303.4650P}
{Patel}, P., {Abdalla}, F.~B., {Bacon}, D.~J., {et~al.} 2013, ArXiv: 1303.4650

\bibitem[{{Patnaik} {et~al.}(2001){Patnaik}, {Menten}, {Porcas}, \&
  {Kemball}}]{2001ASPC..237...99P}
{Patnaik}, A.~R., {Menten}, K.~M., {Porcas}, R.~W., \& {Kemball}, A.~J. 2001,
  in Astronomical Society of the Pacific Conference Series, Vol. 237,
  Gravitational Lensing: Recent Progress and Future Go, ed. T.~G. {Brainerd} \&
  C.~S. {Kochanek}, 99

\bibitem[{{Planck Collaboration} {et~al.}(2013){Planck Collaboration}, {Ade},
  {Aghanim}, {Armitage-Caplan}, {Arnaud}, {Ashdown}, {Atrio-Barandela},
  {Aumont}, {Baccigalupi}, {Banday}, \& et~al.}]{2013arXiv1303.5076P}
{Planck Collaboration}, {Ade}, P.~A.~R., {Aghanim}, N., {et~al.} 2013, ArXiv:
  1303.5076

\bibitem[{{Refsdal}(1964)}]{1964MNRAS.128..307R}
{Refsdal}, S. 1964, \mnras, 128, 307

\bibitem[{{Scalo} \& {Elmegreen}(2004)}]{2004ARA&A..42..275S}
{Scalo}, J. \& {Elmegreen}, B.~G. 2004, \araa, 42, 275

\bibitem[{{Schneider} \& {Sluse}(2013)}]{2013arXiv1306.0901S}
{Schneider}, P. \& {Sluse}, D. 2013, ArXiv: 1306.0901

\bibitem[{{Sereno}(2004)}]{2004PhRvD..69h7501S}
{Sereno}, M. 2004, \prd, 69, 087501

\bibitem[{{Sluse} {et~al.}(2005){Sluse}, {Hutsem{\'e}kers}, {Lamy}, {Cabanac},
  \& {Quintana}}]{2005A&A...433..757S}
{Sluse}, D., {Hutsem{\'e}kers}, D., {Lamy}, H., {Cabanac}, R., \& {Quintana},
  H. 2005, \aap, 433, 757

\bibitem[{{Suyu} {et~al.}(2010){Suyu}, {Marshall}, {Auger}, {Hilbert},
  {Blandford}, {Koopmans}, {Fassnacht}, \& {Treu}}]{2010ApJ...711..201S}
{Suyu}, S.~H., {Marshall}, P.~J., {Auger}, M.~W., {et~al.} 2010, \apj, 711, 201

\bibitem[{{Suyu} {et~al.}(2009){Suyu}, {Marshall}, {Blandford}, {Fassnacht},
  {Koopmans}, {McKean}, \& {Treu}}]{2009ApJ...691..277S}
{Suyu}, S.~H., {Marshall}, P.~J., {Blandford}, R.~D., {et~al.} 2009, \apj, 691,
  277

\bibitem[{{Suyu} {et~al.}(2013){Suyu}, {Treu}, {Hilbert}, {Sonnenfeld},
  {Auger}, {Blandford}, {Collett}, {Courbin}, {Fassnacht}, {Koopmans},
  {Marshall}, {Meylan}, {Spiniello}, \& {Tewes}}]{2013arXiv1306.4732S}
{Suyu}, S.~H., {Treu}, T., {Hilbert}, S., {et~al.} 2013, ArXiv: 1306.4732

\bibitem[{{Tewes} {et~al.}(2013){Tewes}, {Courbin}, {Meylan}, {Kochanek},
  {Eulaers}, {Cantale}, {Mosquera}, {Magain}, {Van Winckel}, {Sluse},
  {Cataldi}, {V{\"o}r{\"o}s}, \& {Dye}}]{2013A&A...556A..22T}
{Tewes}, M., {Courbin}, F., {Meylan}, G., {et~al.} 2013, \aap, 556, A22

\bibitem[{{Treu}(2010)}]{2010ARA&A..48...87T}
{Treu}, T. 2010, \araa, 48, 87

\bibitem[{{Tsupko} \& {Bisnovatyi-Kogan}(2012)}]{2012GrCo...18..117T}
{Tsupko}, O.~Y. \& {Bisnovatyi-Kogan}, G.~S. 2012, Gravitation and Cosmology,
  18, 117

\bibitem[{{Tsupko} \& {Bisnovatyi-Kogan}(2013)}]{2013PhRvD..87l4009T}
{Tsupko}, O.~Y. \& {Bisnovatyi-Kogan}, G.~S. 2013, \prd, 87, 124009

\bibitem[{{Vegetti} \& {Koopmans}(2009)}]{2009MNRAS.392..945V}
{Vegetti}, S. \& {Koopmans}, L.~V.~E. 2009, \mnras, 392, 945

\bibitem[{{Xu} {et~al.}(2012){Xu}, {Mao}, {Cooper}, {Gao}, {Frenk}, {Angulo},
  \& {Helly}}]{2012MNRAS.421.2553X}
{Xu}, D.~D., {Mao}, S., {Cooper}, A.~P., {et~al.} 2012, \mnras, 421, 2553

\bibitem[{{York} {et~al.}(2005){York}, {Jackson}, {Browne}, {Wucknitz}, \&
  {Skelton}}]{2005MNRAS.357..124Y}
{York}, T., {Jackson}, N., {Browne}, I.~W.~A., {Wucknitz}, O., \& {Skelton},
  J.~E. 2005, \mnras, 357, 124

\end{thebibliography}

\end{document}